\documentclass[twocolumn,aps,prd,longbibliography]{revtex4-2}
\usepackage{amsmath,latexsym}
\usepackage{xcolor}
\usepackage[colorlinks=true, urlcolor=blue, linkcolor=blue, citecolor=blue]{hyperref}
\usepackage{etoolbox}
\usepackage{breqn}
\usepackage{multirow}
\usepackage{tipa}
\usepackage{lipsum}
\usepackage{float}

\usepackage{booktabs}
\usepackage{makecell}
\DeclareUnicodeCharacter{0301}{\'{e}}
\makeatletter
\let\cat@comma@active\@empty
\makeatother

\begin{document}

\title{Dual residence time for droplet to coalesce with liquid surface}
\author{Ting-Heng Hsieh$^1$, Wei-Chi Li$^{1,2}$, and Tzay-Ming Hong$^1$\thanks{ming@phys.nthu.edu.tw} }
\thanks{ming@phys.nthu.edu.tw}
\affiliation{$^1$Department of Physics, National Tsing Hua University, Hsinchu, Taiwan 30013, Republic of China\\
$^2$Department of Physics, Emory University, Atlanta, GA 30322}
\date{\today}

\begin{abstract}
When droplets approach a liquid surface, they have a tendency to merge in order to minimize surface energy. However, under certain conditions, they can exhibit a phenomenon called coalescence delay, where they remain separate for tens of milliseconds. This duration is known as the residence time or the non-coalescence time. Surprisingly, under identical parameters and initial conditions, the residence time for water droplets is not a constant value but exhibits dual peaks in its distribution. In this paper, we present the observation of the dual residence times through rigorous statistical analysis and investigate the quantitative variations in residence time by manipulating parameters such as droplet height, radius, and viscosity. Theoretical models and physical arguments are provided to explain their effects, particularly why a large viscosity or/and a small radius is detrimental to the appearance of the longer residence time peak.
\end{abstract}

\maketitle


\section{Introduction}
Triggered by surface tension, examples of coalescence abound in our daily life, e.g., when milk is poured into coffee or the raindrops fall on a puddle \cite{Coalescenceofliquiddrops, Hydrodynamics, planar,Coalescence_of_Drops}. It is relatively less well-known that the liquid droplet in fact does not necessarily merge immediately with the pool after touching down on its surface. First reported by Lord Rayleigh almost one and a quarter centuries ago \cite{first}, this delay in the coalescence  is ascribed to the need for air to flow out of the intermediate region. The ability to trap and create an air film is made possible by the fact that the surfaces of both the droplet and pool are deformable. Otherwise, 
a solid ball dropping on a pool or a droplet plummeting onto a solid surface will only create a splash \cite{DROP_IMPACT_DYNAMICS:Splashing_Spreading_Receding_Bouncing, Drop_Impact_on_a_Solid_Surface, Drop_Splashing_on_a_Dry_Smooth_Surface,Maximun_air_bubble_entrainment_at_liquid_drop_impact}.

As the air is drained out by the weight of the droplet during the non-coalescence, the liquid molecules on both surfaces can ``feel'' each other as soon as their distance becomes short enough to interact via the van der Waals force. A recent simulation study \cite{Droplet_Coalescence_is_Initiated_byThermal_Motion} shows that the coalescence can actually set in earlier if the thermal fluctuation of surfaces is taken into consideration. Previous researchers have managed to prolong the residence time by (1) renewing the film, e.g., via oscillating the pool surface vertically \cite{From_bouncing_to_floating, Lifetime_of_bouncing_droplet, Dynamics_of_a_Bouncing_Droplet_onto_a_Vertically_Vibrated_Interface} or replenishing the air via (2) creating a relative speed between the droplet and the pool \cite{Levitation_of_a_drop_over_a_film_flow, Impact_of_droplets_on_inclined_flowing_liquid_films, Bowling_water_drops_on_water_surface, Surfing_of_drops_on_moving_liquid_liquid_interfaces} and (3) creating a temperature difference between the droplet and pool \cite{Thermal_delay_of_drop_coalescence}. The latter two methods are so effective that they even work on a solid boundary, e.g., for dynamic hydrophobicity \cite{Levitation_of_a_drop_over_a_moving_surface, Aerodynamic_Leidenfrost_effect, Droplet_levitation_over_a_moving_wall_with_a_steady_air_film} and the Leidenfrost effect \cite{Leidenfrost_drops,Leidenfrost_Dynamics,The_leidenfrost_phenomenon_film_boiling_of_liquid_droplets_on_a_flat_plate}.  

\begin{figure*}
	\centering	
	\includegraphics[width=18cm]{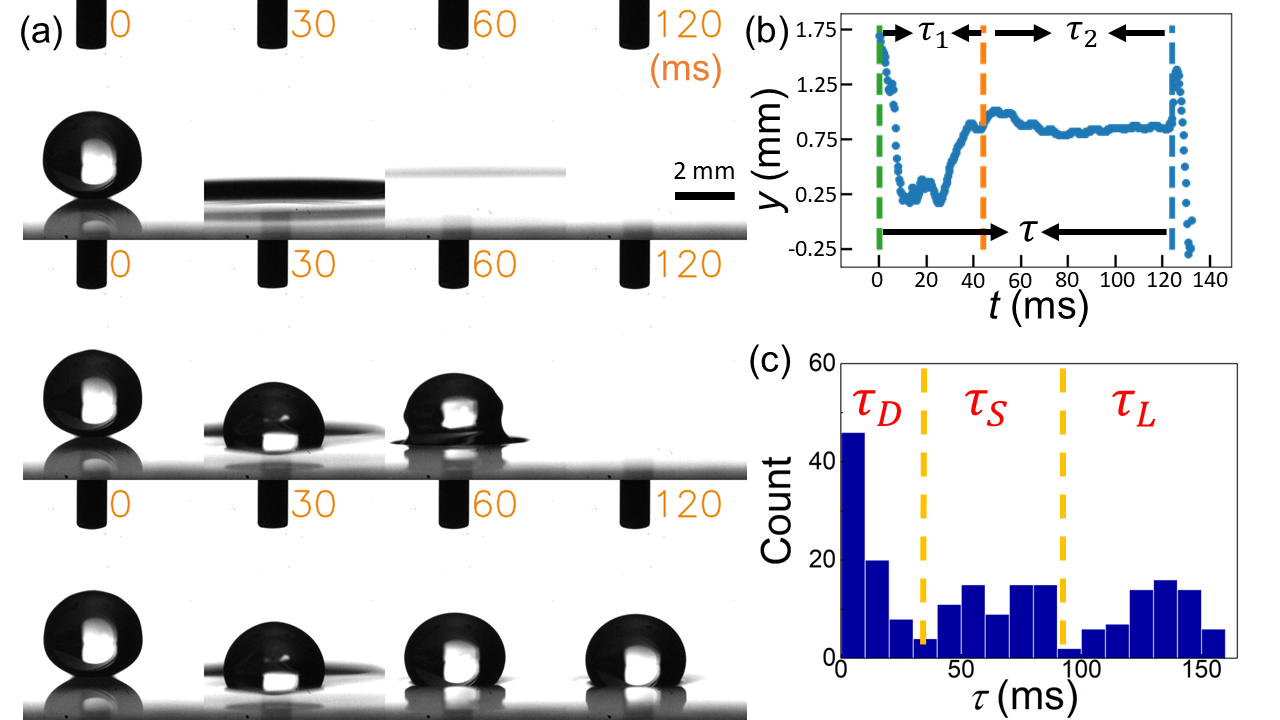}
   \caption{(a) Snap shots of a water droplet with radius $R$=1.79 mm and height $H$=3 mm. Starting from above, three rows of possible outcomes are distinguished by their length of residence time: coalescence, and short and long residence. (b) For the latter two cases, the droplet survives the oscillation stage and remains roughly still until suddenly collapsing. (c) A typical histogram for the residence time.}
    \label{ introduction}		    
\end{figure*}

If released from a medium height $H$, a non-coalescing droplet undergoes four stages: free fall, oscillating vertically, floating, and coalescing, as shown in Fig. \ref{ introduction}(a). The oscillation time, defined as $\tau_1$ in Fig. \ref{ introduction}(b), starts from the emergence of ripple causing by the impact
of droplet on the pool surface and ends at the moment when it ceases to oscillate. The residence time $\tau$, denoting the sum of $\tau_1$ and the remaining time until coalescence $\tau_2$, is approximately 100 ms for the droplet radius $R \approx$ 1.5 mm. 
It is worth mentioning that there exists a critical height $H_c$ beyond which the droplet will simply smash onto the pool and render $\tau=0$. Physically it can be estimated by requiring the kinetic energy $\rho R^3gH_c$ to be comparable to the surface energy $\gamma R^2$. The ratio of these energies is the Weber number.

Klyuzhin {\it et al.} \cite{Persisting} have reported that residual charges on the water droplets can give rise to a second residence time. Encouraged by their nice work,  we set out to explore the effect of  $H$ and $R$ on the dual residence time. Rather than averaging the distribution of $\tau$ as in Ref.\cite{Persisting}, we distinguish different residence times and their corresponding probability by the K-means clustering \cite{means} which is an iterative algorithm capable of objectively determining the most statistically meaningful number of peaks and their corresponding weight from any distribution.
As exemplified in  Fig. \ref{ introduction}(c), it turns out to be the norm rather than the exception that there exist three  residence times, $\tau_D$ for death, and $\tau_S$ and $\tau_L$ for short and long $\tau$. Droplets that coalesce within $\tau_1$ are  designated in the category of $\tau_D$. An interesting observation to bear in mind is that, while $\tau_L\approx 2\tau_S$, their $\tau_1$ roughly equals.

\section{Experimental setup}

\begin{figure}[ht]
	\includegraphics[width=8.5cm]{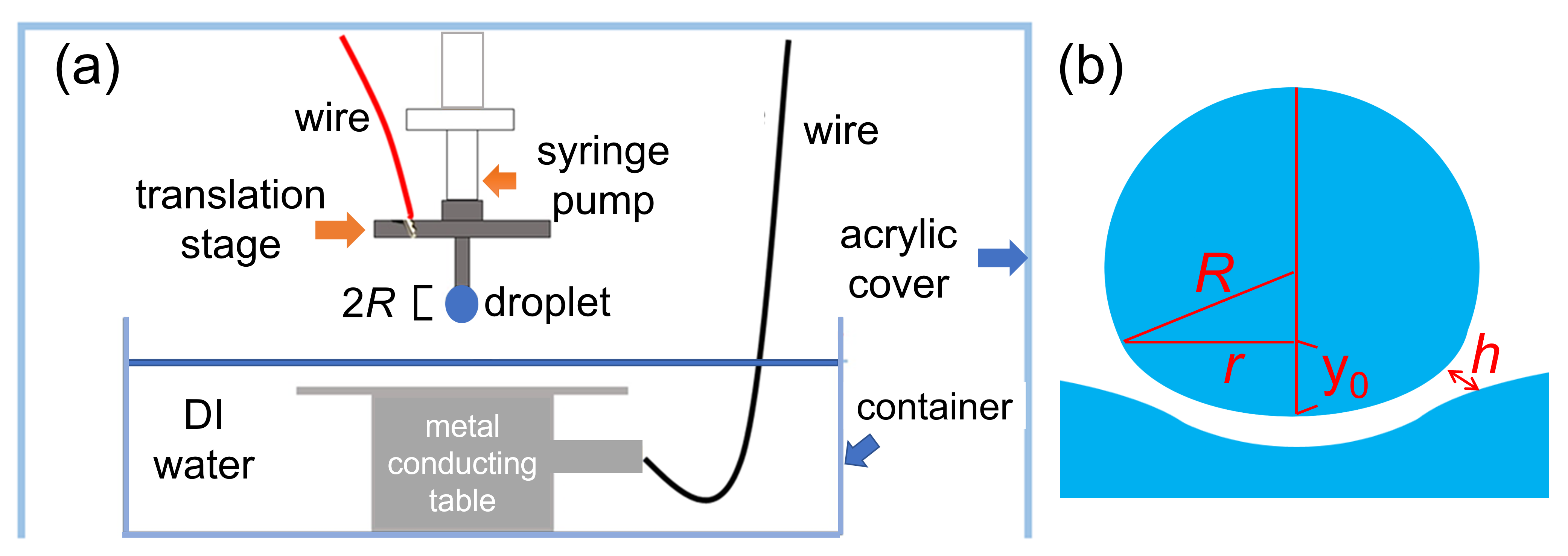}
    \caption{
     (a) Schematic experimental setup for measuring the residence time of water droplets falling on a water pool through the air, (b) Cutaway view of droplet floating on the pool surface with the equilibrium depth $y_0$, the radius of contact area $r$, and film thickness $h$.}
    \label{ setup}		    
\end{figure}

Our goal is to understand when, why, and how the residence time for a droplet to coalesce with the same fluid in the pool should adopt multiple values under seemingly identical experimental conditions (e.g., temperature and humidity) and parameters ($H$ and  $R$). 
Schematic experimental setup can be found in Fig. \ref{ setup}(a). 
The radius of the acrylic container is 15 cm, big enough to ensure the reflected ripple that takes about 1 s to come back will not interfere with the residence time $\sim$ 0.2 s. A syringe pump is used to make sure the output speed of the droplet is negligible. The metallic needle of the syringe is grounded and replaceable that comes in five sizes. The grounded conducting table is to avoid an accumulation of static charge in the pool. Adjusted by a translation stage with precision $\le$ 0.01 mm, the height $H$ of the droplet measures the displacement of its center of mass that ranges from 1 mm to 10 mm. The whole process of non-coalescence is captured by a high-speed camera (Phantom Miro-ex4) with frame rate of 1000 fps. An acrylic cover is employed throughout the experiment to prevent disturbance from the wind. 

\section{Theory}


From dimensional analysis, we can model the motion of droplet during $\tau_1$ stage by an underdamped harmonic oscillator\cite{Contact_time_of_a_bouncing_drop, A_qusai_static_model_for_drop_impact, Viscous_bouncing} where the surface tension coefficient $\gamma$ plays the role of spring constant:
\begin{equation}
    \rho R^{3}\ddot{y}+\alpha\dot{y}+ \gamma y =0.
\end{equation}
Setting $y=A\cdot\exp{(i \omega t)}$ gives
\begin{equation}
    \omega=\frac{ i(\alpha/R) \pm \sqrt{\rho R\gamma-(\alpha/R)^{2}}}{2\rho R^{2}}.
\end{equation}
The term inside the square root represents the periodic oscillation and the other term represents the decrease in amplitude. We assume that $\tau_1$ ends when the amplitude is smaller than some minimal discernible amplitude. Incorporating the information that the initial amplitude equals $y_1-y_0$ where the maximum and equilibrium depths respectively obey $\rho R^3 gH\approx\gamma y_1^2$ and $\rho R^3 g\approx \gamma y_0$ with $g=9.8\ {\rm m/s^2}$ leads to 
\begin{equation}
    \tau_1= \frac{\rho R^{3}}{\alpha}\ln\big{(}\sqrt{\frac{\rho R^3 gH}{\gamma}}-\frac{\rho R^3 g}{\gamma}\big{)}.
\label{tau1}
\end{equation}
Note that in addition to the obvious source of dissipation from viscosity $\mu$, the creation of ripples also takes away some mechanical energy from the droplet. Therefore,  the damping coefficient should include two terms as $\alpha= \mu R+ \sqrt{\gamma \rho R^3}$.

Now let's examine the residence time from the perspective of how long the film can be sustained.  By use of the Pythagorean theorem, the radius  $r$ of contact area in Fig. \ref{ setup}(b) can be derived by the knowledge of $y_0$ as
 \begin{equation}
     r=\sqrt{ \frac{ \rho R^{4} g} {\gamma} -\frac{\rho^{2}R^{6} g^{2}}{\gamma^{2}}}.
 \label{smallr}
 \end{equation}
 Note that the Bond number Bo=$\rho g R^2/\gamma\approx$ 0.15 for $R$=1.1 mm which implies that the effect of gravity is small but not negligible compared to the surface tension.  Intuitively, we expect the gravity will squeeze the droplet and lead to a roughly flat bottom surface, as depicted in Fig. \ref{ setup}(b). The resulting increase in $r$ is equivalent to multiplying Eq. (\ref{smallr}) by a coefficient of (1+$\mathcal{O}$(Bo)). But since the original Eq. (\ref{smallr}), $r=r_0 \sqrt{1-{\rm Bo}}$, can be treated as $r_0$ plus a leading term of $\mathcal{O}$(Bo). This is of the same order of magnitude as the correction from (1+$\mathcal{O}$(Bo)). Therefore, we expect the effect of gravity or the deformation of the droplet not to alter the following theoretic predictions in a qualitative way.

Drained by the difference  $\Delta P\equiv P_f -P_0$  between the pressure inside the air film $P_f$ and the ambient $P_0$, 
the air has to pass through an opening whose area roughly equals $2 \pi r h$. 
It will be estimated in Sec. IV(E) that the Reynolds number roughly equals 0.5, i.e., the viscous stress is not negligible compared with the inertia.
For the sake of simplicity, let's employ Bernoulli's principle for the time being to obtain an analytic expression for the relation between the drift velocity $v$ and  $\Delta P$. 
\begin{equation}
\Delta P \approx  \rho R^3 g/r^{2} = \rho_a v^{2}
\label{bernoulli}
\end{equation}
where $\rho_a$ is the mass density of air. The continuity equation dictates that
\begin{eqnarray}
    -\frac{d (r^2 h)}{d t} \sim rh \cdot v.
\label{comti}
\end{eqnarray}
where the film volume has been approximated by the product of its base area $ \pi r^2$ and thickness $h$.
Since $h \ll y_0$, $r$ can be treated as a constant when solving Eq.  (\ref{comti}) to obtain
\begin{equation}
    \tau \approx \sqrt{\frac{\rho_a}{\rho g}} \big{(}\frac{\rho R^{2.5}g}{\gamma}-\frac{\rho^{2}R^{4.5}g^{2}}{\gamma^{2}}\big{)} \ln (\frac{h_i}{h_f})
    \label{tau2}
\end{equation}
where $h_i$ and $h_f$ are the initial and final thickness. It will be shown in Appendix B to capture the same $\tau$ vs. $R$ behavior as when viscosity is taken in account.


\section{Experimental results}

Presumably, the dual residence time has to derive from the dichotomy of some experimental property, e.g.,  oblate vs. prolate shape for the droplet, zero vs. nonzero triboelectric effect, inward vs. outward internal flow, and symmetric vs. asymmetric geometry for the film. To sift through these candidates, we shall systematically vary the height, shape, and radius of the droplet, viscosity, and voltage between the syringe and pool to clarify their roles. 
\subsection{Flow field and particle image velocimetry}

\begin{figure}[ht]	
	\includegraphics[width=8.5cm]{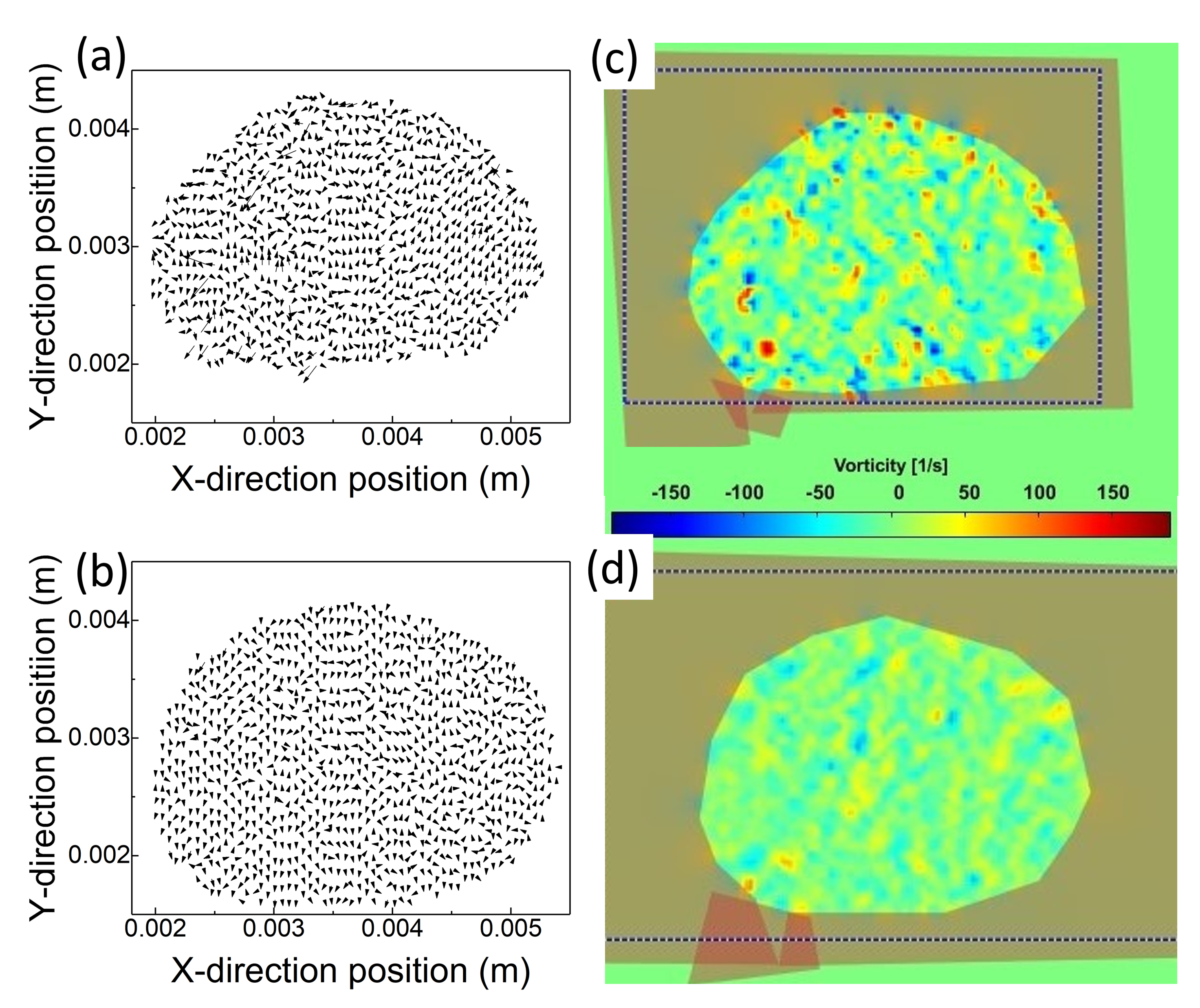}
   \caption{The averaged flow vectors for the first ten millisecond are determined by PIV for (a) $\tau_S$ and (b) $\tau_L$. Both of $\tau_S$ and $\tau_L$ shows no specific direction of flow. The corresponding vorticity in (c) and (d) rules out the possibility of inward vs. outward internal flow. }
    \label{fig:PIV}		    
\end{figure}

Previous research \cite{Bowling_water_drops_on_water_surface} has demonstrated that rotating droplets can exhibit a significantly longer residence time, say, from 0.1 to 1 s for radius $R$= 1.5 mm and height $H$= 6 mm. Since the droplets are controlled to be stationary in our experiment, the knowledge of Ref.\cite{Bowling_water_drops_on_water_surface} opens up a possible explanation for the dual values of residence times, i.e., an inward vs. outward flow pattern. To investigate this scenario, 
we employed Particle Image Velocimetry (PIV) which is composed of a laser of 2.5 W and 532 nm and a convex lens passing through the optical path, illuminating the titanium dioxide particles inside the droplet. The reflected light signal is picked up by a high-speed camera with 2000 fps. Afterward, data are converted to images via PIVlab of the MATLAB toolbox for the analysis of flow direction. We measured the flow field during the first ten millisecond of $\tau_2$ stage for both $\tau_S$ and $\tau_L$. The reason why the $\tau_1$ stage was excluded was that oscillations of the droplet made the measurement of PIV challenging.  
As shown in Fig. \ref{fig:PIV} where the average velocity has been subtracted, there are no  internal circulation for $\tau_S$ and $\tau_L$. This rules out  the pattern of flow field as a source  for the dual residence time.

\subsection{Coalescence point}
\begin{figure}[ht]
	\includegraphics[width=8cm]{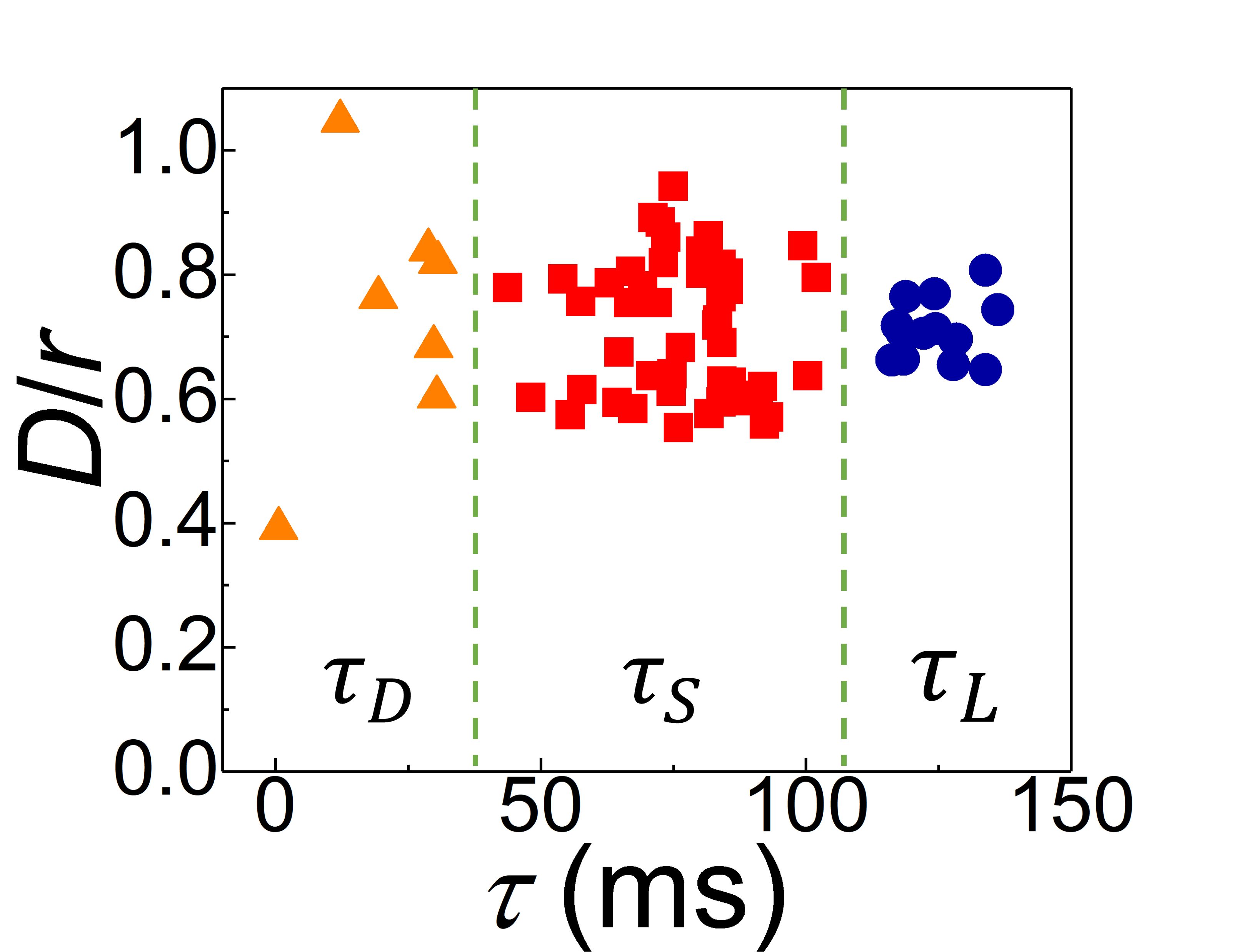}
        \caption{Distance of coalescence $D$, normalized by the film radius  $r$, is plotted against the residence time $\tau$ for $R$= 1.79 mm and $H$=4 mm. Note that droplets always coalesce outside of $r/2$-range for both $\tau_S$ and $\tau_L$.} \label{fig:CP}		    
\end{figure}

Imagine throwing an upside-down bowl on water, it can sink either quickly with a kerplunk sound or slowly with a series of gurgling blubs. The difference lies in how level the bowl rim is with respect to the water surface. 
This led us to an alternative theory for the dual residence time, i.e., the geometric shape of the air film. We recorded the processes both prior to and during coalescence by placing a high-speed camera under the pool. 
 Within the scenario of a sinking bowl, coalescence is expected to start nearer the middle of the droplet for the symmetric or $\tau_L$ case. So, if we define   $D$ as the distance between the point of coalescence to the film center, it ought to be negatively correlated to $\tau$. However, the data in  Fig. \ref{fig:CP} did not confirm this picture, although they do support a negative rather than positive correlation. This failure is partly ascribed to the fact that  $D>r/2$ for both $\tau_S$ and $\tau_L$. There is a simple explanation for this observation:  the air chamber is thicker in the middle than on the edge, according to Ref.\cite{Failure_mechanisms_of_air_entrainment,Air_entrainment_during_impact_of_droplets_on_liquid_surfaces}. 

\subsection{Net charge and external electric field}
Klyuzhin {\it et al.} \cite{Persisting}   have suggested that the electrical charge plays a significant role in shortening the residence time. They observed that the instances of shorter residence time could be erased by grounding the pool. This inspired us to examine a third scenario for the dual residence time, i.e., charged vs. neutral or with a different quantity of charges. The Kelvin water dropper famously demonstrated the existence of and feasibility to generate static charges by passing water through a metal ring on the hose.  We believe the detachment of droplets from the needle is no different. 

\begin{figure}[ht]
	\includegraphics[width=8.5cm]{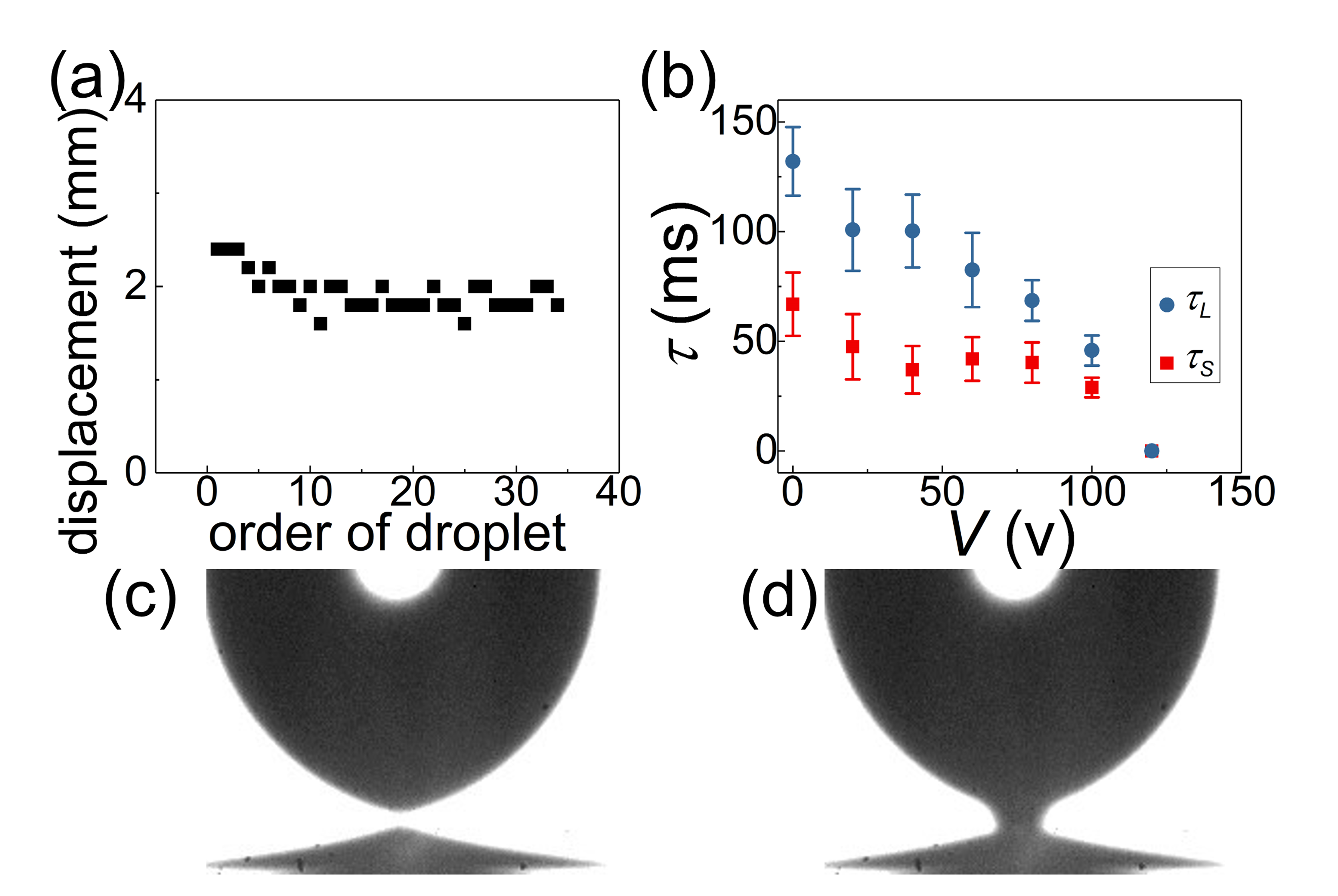}	
    \caption{(a) Horizontal displacement of droplets passing through two parallel metal plates separated by 20 mm and with a voltage difference of 3000 V. The $x$-parameter denotes the order of each trial. 
(b)    Residence time $\tau$ vs. voltage $V$ with $R=1.79$ mm and $H=3.00$ mm.  Both $\tau_S$ and $\tau_L$ drop as $V$ increases. When  $V \ge$ 120 V, all droplets immediately coalesce with the pool due to the formation of the Taylor cone, as evidenced by the snapshots in (c) and (d) taken before and after the development of the water column. To enable better visualization of the Taylor cone, the photos were taken with a much higher voltage at 1000 V.  }
    \label{fig:V_tau}        
\end{figure}

To quantitatively measure possible charges accumulated on the droplets, we employed two parallel vertical metal plates with a voltage difference of 3000 V. After being released from the needle, the droplets pass through the junction of plates and get deflected horizontally with a  displacement that varies with the charges they carry. It turns out that all the droplets appeared to exhibit a similar displacement in Fig. \ref{fig:V_tau}(a), namely, there are no double peaks in their distribution. These droplets are released high above the plates to avoid induction by the electric field. The price we paid was that the droplets would experience the nonuniform field at the plate edges, which complicated the determination of charges. However, since the data show that all droplets share roughly the same displacement, we can at least be sure that they carry similar charges.
 This excludes the static charge as a candidate to explain the dual residence time. 

It is worth mentioning that we also conducted experiments to apply an external voltage $V$ up to 150 V to the needle by connecting it to a DC power supply, while placing a metal plate under the water surface to ground the pool. As depicted in Fig. \ref{fig:V_tau}(b), both $\tau_L$ and $\tau_S$ decrease with $V$. A heuristic argument may be that the Coulomb attraction between the charged droplet and the pool adds more pressure on the film and thus accelerates the drainage of its air.  It is interesting to note that both $\tau_L$ and $\tau_S$ vanish simultaneously at 120 V when a Taylor cone \cite{Taylor,The_Fluid_Dynamics_of_Taylor_Cones,Effect_of_electric_fields_on_the_rest_time_of_coalescing_drops}  emerges from the pool and ``penetrates'' the air film, as shown in Fig. \ref{fig:V_tau}(c, d).

\subsection{Height and shape of droplet}
\begin{figure}[ht]
	\includegraphics[ width=8.5cm]{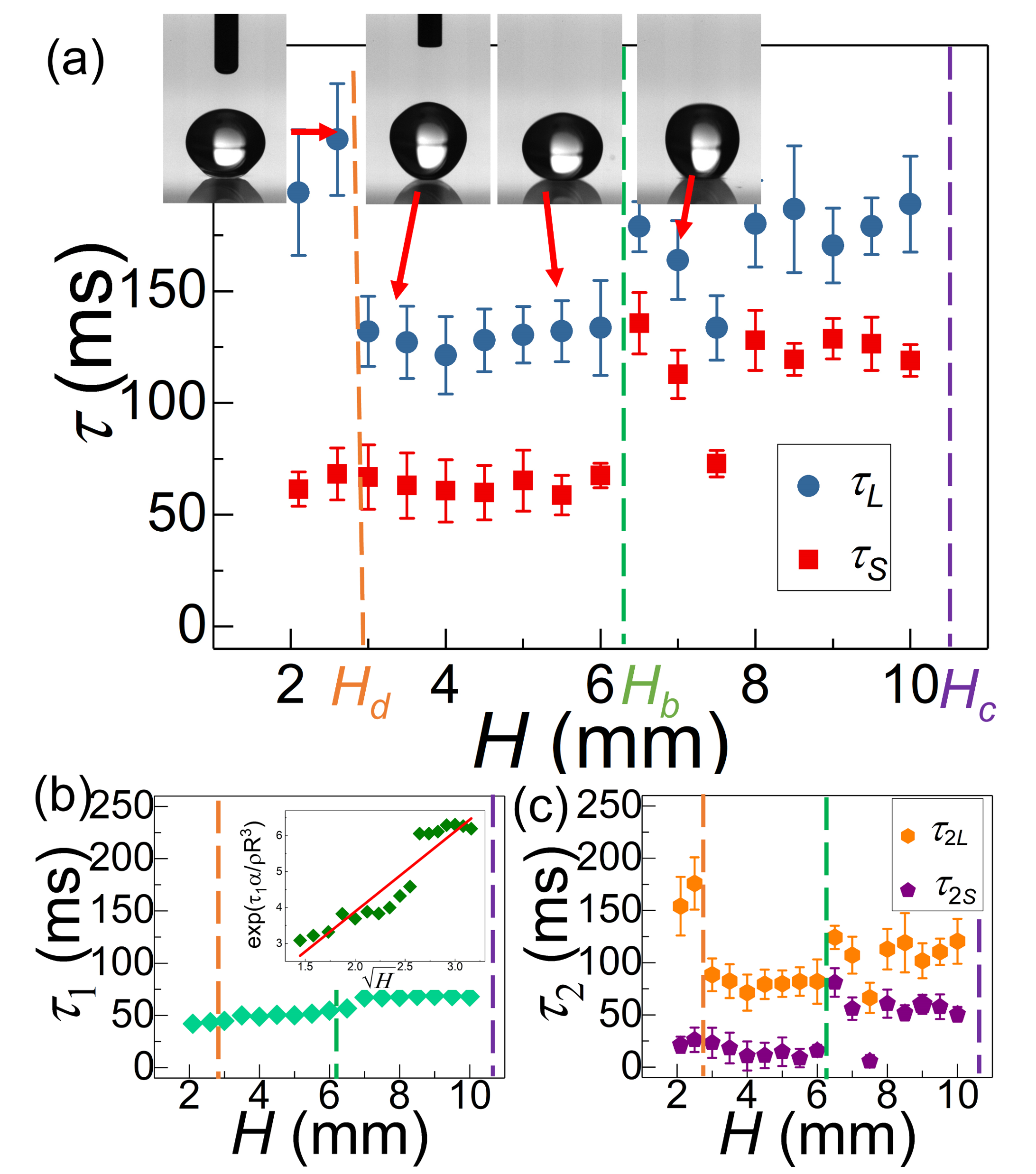}
    \caption{ Residence time $\tau$,  $\tau_1$, and $\tau_2$ are plotted respectively in (a, b, c) as a function of  height for water droplet of radius $R=1.79$ mm in air. Snapshots of droplets upon touching the pool are inserted in (a)  to indicate how its shape alternates between oblate and prolate with $H$.  The straight line in the inset of (b) for $\exp{[\tau_1\alpha/(\rho R^3)]}$ vs. $\sqrt{H}$ is the fitting curve of Eq.  (\ref{tau1}) with an R-squared value of 0.93.}
    \label{ h_resident_time}        
\end{figure}

If the droplet is released at $H=0$, we expect the coalescence to proceed immediately\cite{Noncoalescing_droplet} and $\tau =0$. In contrast, when  $H$ is  high enough to convert a kinetic energy $mgH_c$ that is comparable to the surface energy $4\pi\gamma R^2$, the droplet should either break up or ``penetrate'' the water surface of the pool, i.e., $\tau$ is again 0. Therefore, a naive guess is that $\tau$ ought to follow a convex function for $0\le H\le H_c$.  However, a different scenario was offered in Ref.\cite{Coalescence_of_Drops, Universal_Behavior_Stage} that proposes $\tau$ should decrease monotonically with $H$. Their arguments are that, if $P_f$ increases with $H$, the drainage speed $v$ that results from $\Delta P$ should go up and reduce $\tau$. 

It turns out that the actual $\tau(H)$ deviates from both predictions, as shown in Fig. \ref{ h_resident_time}(a). Three regions can be identified by their distinct behavior. When $H$ is between $H_d$ and $H_b$, $\tau_L$ and $\tau_S$ are roughly independent of $H$. This is very different from the impact of droplet on a flat solid surface where the film thickness was found to vary sensitively with the impact velocity \cite{Maximun_air_bubble_entrainment_at_liquid_drop_impact}.
As $H$ decreases below $H_d$,  $\tau_L (H)$, still a plateau, adopts a larger value discontinuously, while $\tau_S$ remains the same. For $H$ higher than $H_b$, both $\tau_L$ and $\tau_S$ increase continuously by an increment of about 50 ms before $\tau$ becomes zero when $H\ge H_c$.


As soon as the pool surface shows signs of disruption, we start counting $\tau$ for this sandwich structure of droplet, air film, and pool. When detached from the syringe, we intuitively expect the droplet to be deformed and switch from resembling rugby to a discus alternatively during free fall.  
However, two observations rule out this dichotomy as the cause of dual residence time. First, the shape of the droplet  at $t$=0 is indistinguishable for $\tau_L$, $\tau_S$, and even $\tau_D$ in Fig. \ref{ introduction}(a). Second, the middle two photos for $H_d <H<H_b$ in Fig. \ref{ h_resident_time}(a) do exhibit dissimilar shapes, rugby vs. discus-like, but their $\tau_{L,S}$ are roughly the same.

Aided by the high-speed camera, we observe a ring of crest, created by the recoil of the water filament after pinch-off, traveling down the droplet surface for $H< H_d$. Video can be found in SM \cite{sm}. In our opinion, this serves to shove and replenish the air into the film as the crest reaches the bottom of the droplet, which results in a larger value for $\tau_L$. Since $H_d$ corresponds to the free-fall distance within half a vibration period in this scenario, it can be easily calculated as:
\begin{equation}
   H_d\approx (g/2)(\pi\sqrt{m/\gamma})^2\approx \rho R^3g/\gamma
   \label{hd}
\end{equation}
which roughly equals 2.55 mm. This is in agreement with the empirical  value estimated from the position of the orange dash line in Fig. \ref{ h_resident_time}(a). Note that there is another prediction by this picture, i.e., $\tau$ ought to be zero at $H=H_{d}$ because 
the crest will become an exact duplicate of the sharp end at the top of the droplet, except it 
 appears at the bottom and will ``perforate'' rather than help to retain the film. To search for the short interval around $H_d$ where this physics should occur, we refined the increment  of $H$  and indeed found\cite{sm} a tiny window where the distribution of $\tau$ reduces to only containing $\tau_D$.
Since the amplitude of the surface wave is subject to dissipation by the viscosity, its effect on $\tau$ is expected to diminish as $H>H_d$. 

If the syringe is raised even higher, Fig. \ref{ h_resident_time}(a) indicates that new physics enters to lengthen $\tau$ beyond $H_b$. What happens is that the bottom of the droplet is now capable of bouncing above the pool surface and allowing ample fresh air to enter the film.  Based on this picture, $H_b$ can be estimated by setting $y_1 -y_0= y_0$:  
\begin{equation}
 H_b \approx 4 \rho R^3g /\gamma
  \label{hb}
\end{equation}
which is about 6 mm for $R$=1.79 mm, consistent with Fig. \ref{ h_resident_time}(a).

The weak dependence on $H$ for $\tau_1$ in Fig. \ref{ h_resident_time}(b) is consistent with the theoretical result in Eq.  (\ref{tau1}). The reason why the data points for $H>H_b$ seem to deviate from this prediction is that the maximum height after bouncing is roughly the same for different $H$.   For the sake of comparison, Fig. \ref{ h_resident_time}(c) is obtained by subtracting Fig. \ref{ h_resident_time}(b) from (a).

\subsection{Drop radius}

\begin{figure}[ht]
	\includegraphics[ width=8.5cm]{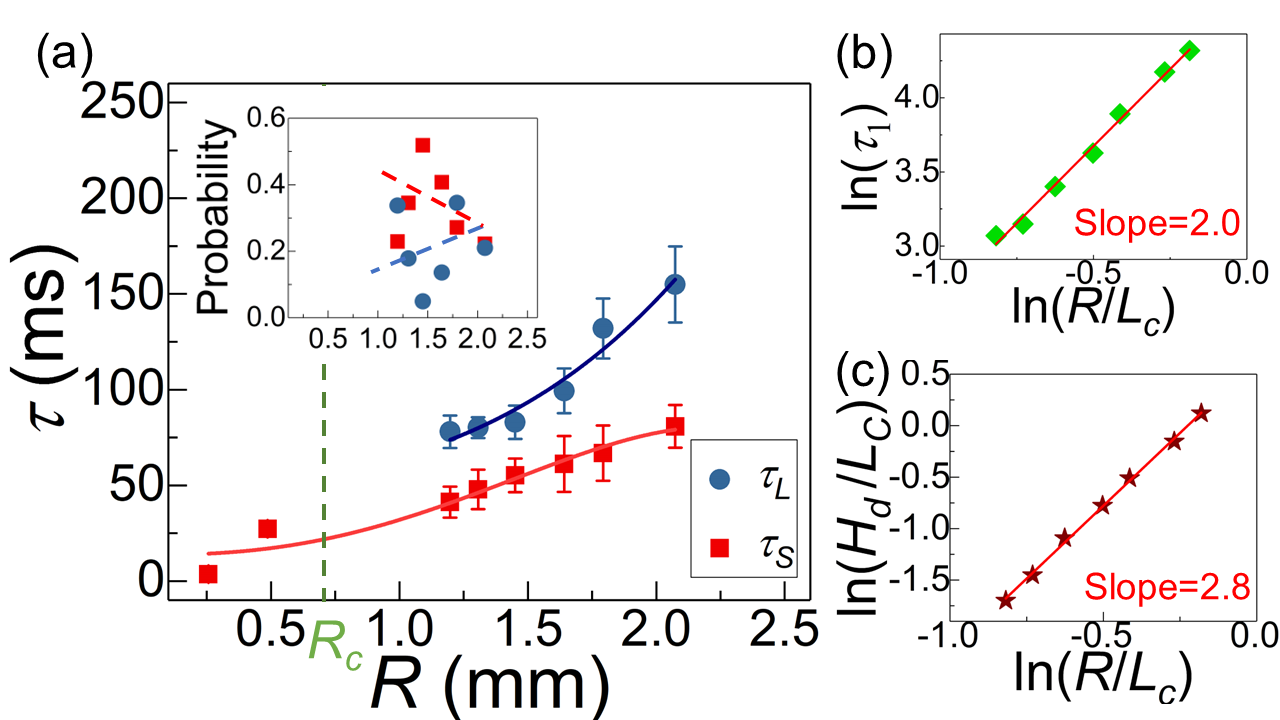}

    \caption{ (a) $\tau$ vs. $R$  for $H_d<H<H_b$. Solid lines are fitting curves based on Eq.  (\ref{tau2}).  Note that dual  $\tau$ values only occur when $R\ge R_c$, during which the probability weighting gradually shifts from  $\tau_L$ to $\tau_S$ as $R\rightarrow R_c$ in the inset.   (b) $\tau_1$ varies as $R^2$  for $H_d<H<H_b$. (c) $H_d$ is found to be proportional to $R^{2.8}$, consistent with Eq.  (\ref{hd}). Solid lines in (b, c) are linear fitting curves. The capillary length $L_c\equiv\sqrt{ \gamma/\rho g}$ is used to render the parameters $R$ and $H_d$  dimensionless. }
    \label{ R_tau}        
\end{figure}

By adding minute amounts of surfactant,  Amarouchene {\it et al.} \cite{Noncoalescing_droplet} studied the maximum height below which non-coalescence can occur. Perhaps due to the presence of surfactant, it did not occur to them that the data might support the dual residence time. After averaging, $\tau $ was claimed to vary linearly with $R$ in accordance with the prediction of lubrication theory that is based on the assumption that the viscous stress dominates the inertia. However, the Reynolds number  ${\rho_a v h}/{\mu_a}\approx 0.5$ for the drainage of air film in our system does not meet such a criterion,  where $\rho_a$ and $\mu_a$ denote the mass density and viscosity of air, drift velocity $v\approx$ 3 m/s, and film thickness $h\approx 10^{-6}$ m \cite{Noncoalescing_droplet}. Aside from the fact that we did not use any surfactant, our conclusion that there exist dual residence times is also different from that of Ref.\cite{Noncoalescing_droplet}. This justifies the need for us to attempt alternative models in Sec. III which targets both $\tau$ and $\tau_1$. Regarding $\tau$,  the solid lines in Fig. \ref{ R_tau}(a) are
the prediction of Eq.  (\ref{tau2}), which
enjoy excellent R-square values of 0.93 and 0.90 for $\tau_L$ and $\tau_S$.

Sharp-eyed readers may notice the scarcity of data for small $R$ in  Fig. \ref{ R_tau}(a). It is constrained by technical difficulty, i.e., to successfully pump fluid out of a needle of radius $a$ involves overcoming the surface tension and thus requires a pressure $\approx\gamma a/a^2$.
The huge pressure to produce really fine droplets thus may cause leaking from the joints. We came up with a way to circumvent the problem, i.e., utilizing the phenomenon of coalescence cascade or partial coalescence  \cite{partialcoalescence, The_coalescence_cascade_of_a_drop, The_mechanism_of_partial_coalescence_of_liquid_drops_at_liquid/liquid_interfaces}. Each generation of daughter droplets performs a few rebounds before partially coalescing, which is when we measure its radius, height, and $\tau$, before the next generation repeats the whole process. Although the trend of these data is equally consistent with our prediction as those produced by the needle in Fig. \ref{ R_tau}(a), we were surprised to find them all single-valued, i.e.,   small radius seems to disfavor the appearance of the dual residence time. Via monitoring the change rate of probability with $R$ in the inset of Fig. \ref{ R_tau}(a), we conclude that it was $\tau_L$ that gets suppressed at small-$R$.

Now let's move on to the second prediction targeting $\tau_1$ in Sec. III, i.e., Eq.  (\ref{tau1}). 
Its anticipation for $\tau_1$ to increase with $R$ is intuitive because more kinetic energy released by a bigger droplet from the same $H$ should take longer for the pool to dissipate. The fitting line in  Fig. \ref{ R_tau}(b) shows that $\tau_1$ is roughly proportional to $R^{1.5}$. We checked that this is consistent with Eq.  (\ref{tau1}) due to two numerical coincidences. First, the logarithmic term in  Eq.  (\ref{tau1}) is insensitive to $R$  since our choice of $H_d <H<H_b$ is close to  $H=H_b/2$ where its derivative equals zero. Second, the inertia term is bigger than the viscous term in $\alpha$. 

Finally, we would like to examine the $R$-dependence of Eq.  (\ref{hd}) for $H_d$. It was not included in Sec. III because we did not foresee its existence until we actually performed the experiments. As shown in Fig. \ref{ R_tau}(c), the prediction of Eq.  (\ref{hd}) based on simple physical arguments is a success with an R-square of 0.98. 

\subsection{Viscosity}
\begin{figure}[ht]
	\includegraphics[ width=8.5cm]{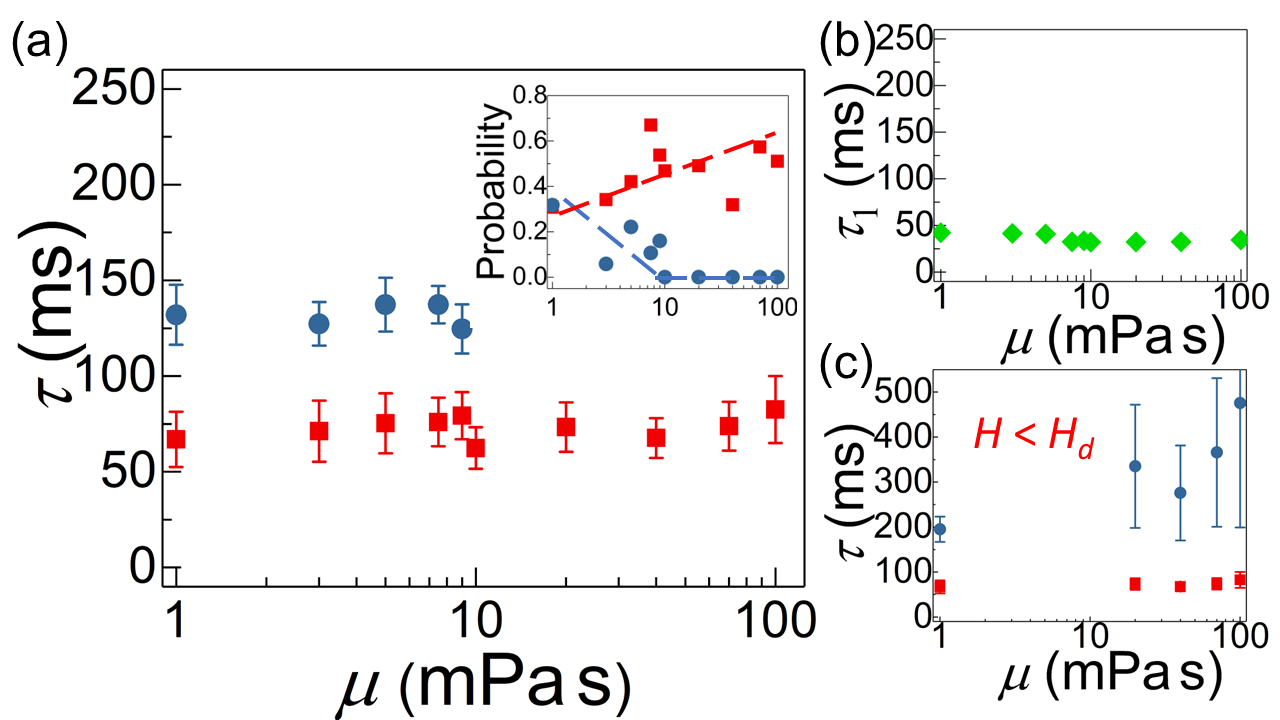}
    \caption{(a) Residence time $\tau$, roughly independent of viscosity $\mu$ for $R$=1.79 mm and $H_d<H$=4 mm$<H_b$, returns to single value when $\mu$ $\ge$ 10 mPa$\,$s. The trend of probability shift in the inset indicates that it is $\tau_L$ that gets suppressed at large  $\mu$. (b) $\tau_1$ decreases as $\mu$ increases, as predicted by Eq.  (\ref{tau1}). (c) For $H < H_d$, $\tau_S$ remains insensitive to $\mu$ as (a) for $R$=1.79 mm, while $\tau_L$ increases with $\mu$ before vanishing at $\mu$=400 mPa$\,$s, consistent with our expectations in Sec. IV(F).  }
    \label{ Vis_tau}        
\end{figure}

Intuitively,  $\tau_1$ should be sensitive to the viscosity $\mu$ because the droplet is oscillating on the pool surface. How about $\tau_2$? As explained at the beginning of this section, one possible explanation for the dual residence time is an inward/outward internal flow in the droplet that replenishes/depletes the air in the film. If the flow indeed exists, we would expect $\mu$ to also affect  $\tau_2$. Unfortunately, this is ruled out by our employment of Particle Image Velocimetry.
How does $\mu$ help clarify the plausibility of our final candidate, i.e., the uniformity or the lack of it for the film thickness? 
A droplet with a large viscosity is slow to deform due to its small velocity gradient. Therefore, it behaves like a lever with a large bending modulus in the time span of $\tau_2$, about tens of millisecond.
In other words, once tilted, the droplet will remain lopsided and create a large opening that allows the trapped air to rush out easily, consistent with the disappearance of  $\tau_L$ when $\mu \ge$ 10 mPa$\,$s in Fig. \ref{ Vis_tau}(a).

To investigate how viscosity affects $\tau$ and its distribution experimentally, we modify the ratio of glycerol-water mixture. The viscosity of glycerol is three orders larger than water, but its surface tension coefficient and mass density are only 10$\%$ less and 25$\%$ higher, respectively. Note that it is the mass $m$, not volume, of the droplet that is determined by the syringe size due to the capillary force. So, when more glycerol is added, the drop size $R$ shrinks, and the film pressure $mg/(\pi r^2)$ depleting the air goes up. Along this line of argument, $\tau_2$ should be shortened as $\mu$ increases. It is at odds with Fig. \ref{ Vis_tau}(a) which shows that $\tau_S$ and $\tau_L$ are roughly independent of $\mu$, whereas  $\tau_1$ decreases in Fig. \ref{ Vis_tau}(b). This discrepancy can be explained by considering the deformation of the droplet around the air outlet, similar to the wind blowing open a window left ajar. As the air drains out from the film, the outlet with a lower viscosity is more susceptible to the widening force inflicted by the air current, which quickens the drainage process and results in a lower $\tau_2$. How about $\tau_1$? According to Eq. (\ref{tau1}), a higher viscosity hastens the energy dissipation and results in a shorter $\tau_1$, which is consistent with Fig. \ref{ Vis_tau}(b). Combined with the opposing trend of $\tau_2$, Fig. \ref{ Vis_tau}(a) thus becomes featureless. 

In Sec. IV(D), we ascribed the enhancement of $\tau_L$ for $H<H_d$ to the replenishment of air in the film by surface waves  triggered by the recoil of liquid filament at pinch-off. Intuitively, we expect this effect to be diminished as the filament becomes thinner and contains less volume if $\mu$ increases. The enhanced dissipation should also hinder the propagation of surface waves. However, $\tau$ turns out to be lengthened in Fig. \ref{ Vis_tau}(c). 
A back-of-the-envelope calculation gives the viscosity length $\mu^2/\rho \gamma \le $ 0.1 mm for $\mu$ as large as 100 mPa$\,$s. The fact that this length is much shorter than the droplet radius $R$=1.79 mm means that the surface waves have no trouble reaching the bottom of the droplet. What we missed in our initial picture is the opposite effect after that, i.e.,  the surface waves will push out some air when they leave the film. The lower crest associated with a larger $\mu$ now becomes less harmful to the sustainment of the film. 



\begin{figure}[ht]
	\includegraphics[width=8.6cm]{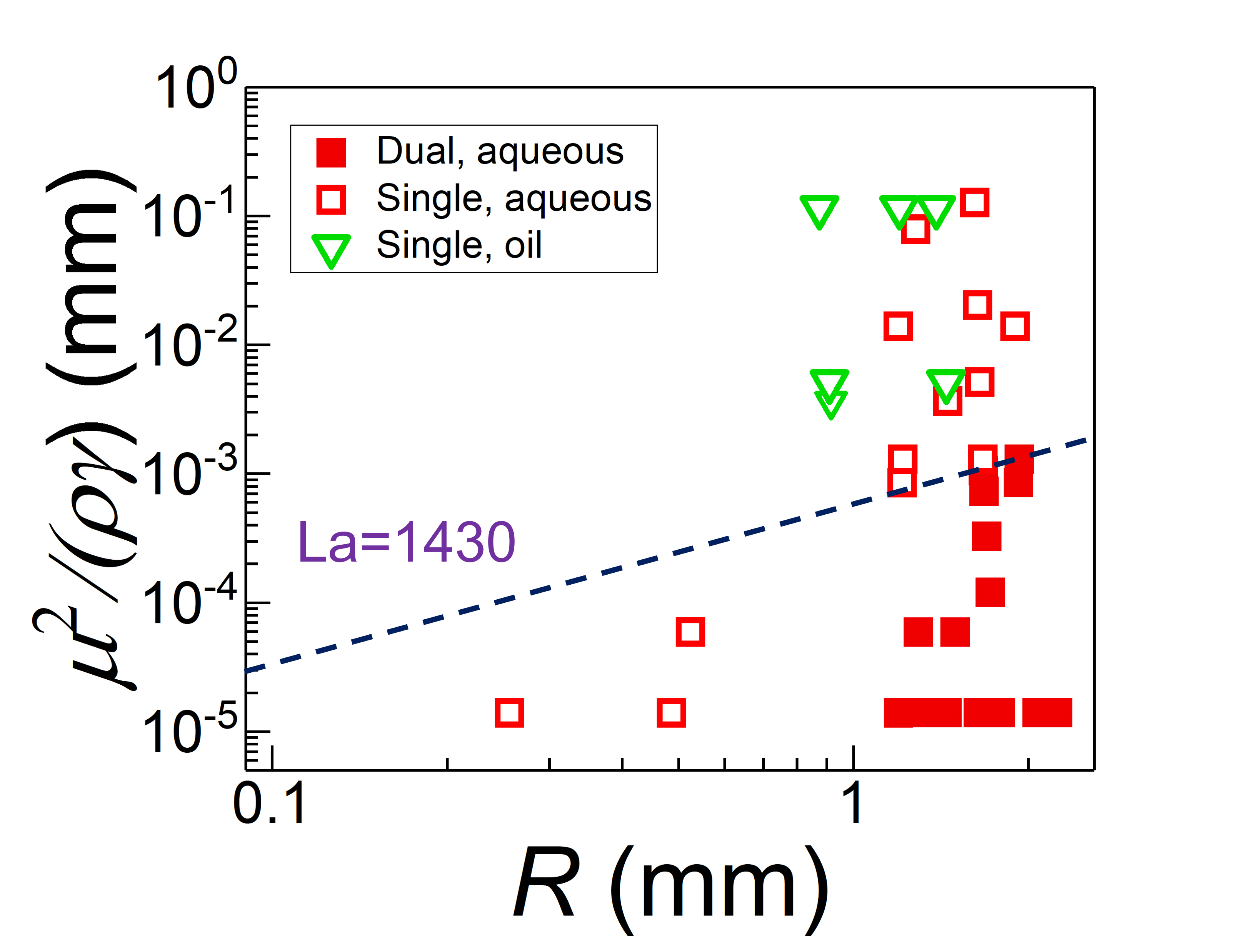}
	
    \caption{Data of different liquid mixtures are compiled to demonstrate that those exhibiting dual residence time and denoted by solid symbols, such as water, glycerin, ethanol, and silicone oil, tend to fall below the blue dashed line, separated from their single-valued counterparts in hollow symbols above. The three small-$R$ points that defy such a rule are created via the coalescence cascade, instead of a syringe like the rest of the samples. }
    \label{ diagram}        
\end{figure} 

Based on the findings in Figs. \ref{ R_tau}(a) and \ref{ Vis_tau}(a), it is tantalizing to assume that there exists a dimensionless threshold that combines $R$ and $\mu$ and beyond which the residence time becomes single-valued. The most obvious candidate is the Laplace number, ${\rm La}=\rho\gamma R/\mu^2$, that  
characterizes the free surface in fluid dynamics.  
By mixing DI water, glycerin, and ethanol, we can prepare samples with different parameters to obtain Fig. \ref{ diagram} that pins the threshold at ${\rm La}=1430$. The only data that are not consistent with this theory are the three points with the smallest $R$ obtained by the coalescence cascade described in Sec. IV(E). We suspect the discrepancy has to do with the pool not completely calming down from the generation of daughter droplets when the latter touchdown. 

Noted that there is no solid triangles in Fig. \ref{ diagram}. For those solid triangles, the data are for liquid alkane, which is much more volatile than the silicone oil denoted by hollow triangles. For the record, their vapor pressure at room temperature is 57.9 and 0.6 KPa, respectively. What happened was that no alkane droplet could survive the oscillating or $\tau_1$-stage. In other words, no nonzero residence time, either single or dual, was observed. We suspect the profuse alkane vapor above the pool contributes to and expedites the formation of liquid bridge (between the droplet and pool surface), which is known to be the precursor of coalescence\cite{Droplet_Coalescence_is_Initiated_byThermal_Motion}. 


\subsection{Symmetric vs asymmetric air film}
\begin{figure}[ht]
	\includegraphics[width=8.6cm]{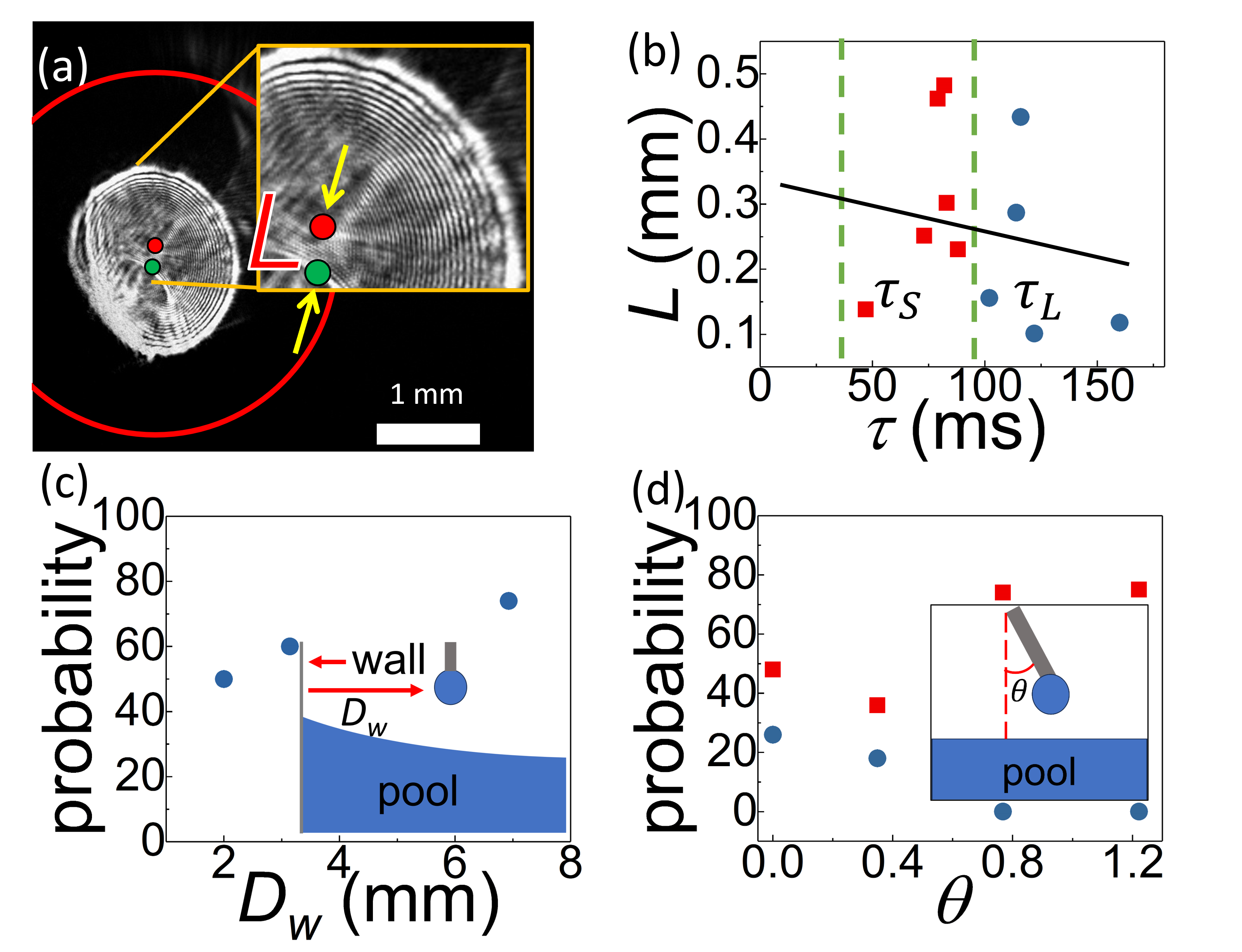}
	
    \caption{(a) A snapshot of the interference pattern during the $\tau_2$ stage for $R$= 1.75 mm and $H$= 3.00 mm. The perimeter of the film is highlighted by the red line whose center deviates from that of the interference pattern by a distance $L$. (b) A negative correlation was observed for $L$ vs $\tau$.  (c) The probability of $\tau_L$ is found to decrease as the droplet approaches the wall. (d) The probability of $\tau_L$ goes to zero when $\theta\ge$ 0.7.  }
    \label{interference}        
\end{figure}

In Sec. IV(F), we conclude that the short residence time is likely caused by the tilt of the droplet. Three efforts were made to provide direct and indirect evidence. Firstly, we tried to deduce the profile of film by interference patterns. Due to the constant leakage of air, the profile is not static so that  a high-speed camera of 8000 fps equipped with  a microscopic lens must be called into action.  Other than that, the employment of a beam splitter while aiming the laser  with wavelength 650nm  from below the pool was standard\cite{Failure_mechanisms_of_air_entrainment, Maximal_Air_Bubble_Entrainment_at_Liquid-Drop_Impact, Spontaneous_self-dislodging_of_freezing_water_droplets_and_the_role_of_wettability, Leidenfrost_droplet_trampolining}. 
Note that the pattern  in Fig. \ref{interference}(a) only covered partially the base area near the drop center. By counting the number of bright fringes, we know that the largest thickness occurs near the middle of the film and is about  10 $\mu$m, while the opening in the perimeter  is no more than 1 $\mu$m. This decrease in the film thickness as we move away from the center is consistent with the observation in Sec. IV(b) that coalescence always happens near the outskirt. 
We measured the distance  $L$ between the centers of the droplet and interference pattern and tried to see whether it could help us distinguish the  $\tau_L$ and $\tau_S$ cases. The negative correlation between $L$ and $\tau$ revealed by Fig. \ref{interference}(b) was consistent with our expectation that an asymmetric film would more likely be associated with a large $L$. The scatteredness of data is not surprising because the profile of flim is subject to the influence of air flow which contains randomness in its spatial distribution of speed.  

So far, we were limited to pursuing consequences passively from the asymmetry which occurred apparently by chance. Why don't we turn the tables and try to create a tilt film proactively? After some thoughts, we came up with two experiments by (1)  making use of the tilt water surface near the wall of the pool  due to wetting and (2)  tilting the needle that connects to the syringe. The results of both attempts in Fig. \ref{interference}(c, d) match our expectations and, therefore, are in support of our theory. As the droplet moves closer to the wall, i.e., when the distance  $D_w$ decreases in Fig. \ref{interference}(c), the degree of tilt is intensified which suppresses the probability of $\tau_L$. In the meanwhile, increasing the tilt angle $\theta$ of the needle in Fig. \ref{interference}(d) gives a similar effect.

\section{conclusion and discussions}
In conclusion, we have investigated the distribution of non-coalescence time for droplets falling on a pool of the same liquid. A strict statistical algorithm helps us establish that there exist three distinct peaks for droplets under identical initial conditions and controlled parameters. The residence time can be divided into two parts: the oscillating time $\tau_1$ to dissipate off most of the kinetic energy and the remaining time $\tau_2$ when the droplet hovers stably over the pool surface. The drainage of air in the film is believed to take place throughout $\tau$. We observed that the height $H$ affects not only the impact velocity but also the shape of the droplet upon contact with the pool. 
But, the existence and value of dual residence time are not sensitive to both factors, except for $H\le H_d$ and $H\ge H_b$ when the surface wave created by pinch-off and rebound of droplet respectively come into play to lengthen $\tau$. We managed to derive analytic expressions for the important physical properties, such as $\tau_1$, $\tau$, $H_d$, and $H_b$, which match well with the experimental results. When theory failed, we did our best to provide heuristic physical arguments to explain at least the qualitative behavior, e.g., why a large $\mu$ or small $R$ suppresses the appearance of $\tau_L$. 

After excluding possible origins, such as oblate vs. prolate shape for the droplet, with vs. without the triboelectric effect, and inward vs. outward internal flow, we attribute the dual residence time to how tilting the droplet is. This is the only scenario that is consistent with all our experimental findings. It is inspired by the ordinary experience of throwing a bowl upside-down on water. We expect two outcomes: the bowl can sink either quickly with a kerplunk sound or slowly with a series of gurgling blubs, depending on how level the bowl rim is with respect to the water surface. 
Similar to the experience from pinching the garden hose, it is thinkable that the side of the film with a smaller opening will exert more force on the droplet to push it back into a more balanced position. This feedback mechanism makes sure that the symmetric or $\tau_L$ case is not a singular event, i.e., valid within a range of small tilt angles.

To test how general our conclusions are and to gain more insights, it will be interesting and recommendable to try different combinations of fluid and medium, i.e., air bubble and DI water where the film composes of water rather than air, and immiscible liquids such as oil and DI water.
Potential applications of our study include (1) the food industry, say, to create distinctive flavor by mixing different oil, (2) the manufacturing, e.g., the fusing of molten particles to form a continuous film, as in welding and brazing, emulsion and coating, and (3) the material science of how microvoids link together at the grain boundary that may lead to the fracture of metallic alloys. 

\section*{Acknowledgement}
We are grateful to Jow-Tsong Shy, Nan-Jung Hsu, and Chia-Wei Deng for technical assistance in the early stage of this project, and to the Ministry of Science and Technology in Taiwan  for financial support under Grants No. 111-2112-M007-025 and 112-2112-M007-015.

\appendix
\section{K-MEANS METHOD}
To distinguish the number and position of peaks as well as their boundaries in a histogram such as Fig. 1 in the main text, we introduced the K-means clustering which is a method of vector quantization, originally from signal processing. The algorithm involves three main steps: First, select the most likely number of peaks in the distribution, say $K$, and then randomly place $K$ points in the data and determine their Voronoi cells. Second, relocate the $K$ points to the centroid of their residing clusters. Finally, repeat the above processes until the algorithm converges. Presumably, a larger $K$ will enable a smaller variance,
\begin{equation}
    \sum_{i=1}^K \sum_{x \in S_i} |x-S_i|^2,
\end{equation}
which $S_i$ represents the centroid of each cluster and $x$ denotes the data points. But we need a balance between the goodness of fit and the simplicity of the model. The trade-off can be objectively determined by calculating the silhouette coefficient which can help us select the best suitable $K$. 

\section{EFFECT OF MEDIUM VISCOSITY FOR $\tau$}
Since the Reynolds number is estimated to be 0.5 in Sec. IV(E), the viscous pressure cannot be neglected. A quick way to modify Bernoulli's principle in Eq. (\ref{bernoulli}) is to deduct the heat from  the gravitational energy before converting it to the kinetic energy of the draining air:
\begin{equation}
 \rho R^3 g h-\mu_a v^2 h \tau\approx \rho_a r^2 h v^2.
\end{equation}
where  the form of $\mu v^2 h$ can be argued from dimensional analysis.
Solving $v$ and plugging in the continuity equation of Eq.  (\ref{comti}) give
\begin{equation}
\tau \approx \bigg{(}\frac{R}{\gamma}-\frac{\rho R^3g}{\gamma^2}\bigg{)}\bigg{(}\mu_a+\sqrt{\mu_a^2  +\frac{\rho_a \rho g R^3}{ \big{(}\ln(\frac{h_i}{h_f})\big{)}^2}}\bigg{)} \big{(}\ln(\frac{h_i}{h_f})\big{)}^2
\label{RNF}
\end{equation}
which reproduces Eq.  (\ref{tau2}) if $\mu_a$ is set to be zero.
Fitting Fig. \ref{ R_tau}(a) by Eq.  (\ref{RNF}) gives R-Squared values of 0.91 and 095 for $\tau_L$ and $\tau_S$, as opposed to 0.93 and 0.90 by Eq.  (\ref{tau2}).

\end{document}